\documentclass[useAMS,usenatbib]{mn2e}
 \usepackage{epsfig}

\title[Fossil AGN jets as ultra high energy particle accelerators]{Fossil AGN jets as ultra high energy particle accelerators}

\author[Gregory Benford \& R.J. Protheroe]{Gregory Benford$^{1}$\thanks{E-mail: G.Benford@uci.edu}, R. J. Protheroe$^{2}$\thanks{E-mail: rprother@physics.adelaide.edu.au}\\
$^{1}$Department of Physics and Astronomy, University of California, Irvine, CA 92697-4575, USA\\
$^{2}$Department of Physics, School of Chemistry \& Physics, University of Adelaide, Adelaide, SA 5005, Australia}
\begin{document}
 
 
 
\maketitle
 
\label{firstpage}
 
\begin{abstract}
Remnants of AGN jets and their surrounding cocoons leave colossal
magnetohydrodynamic (MHD) fossil structures storing total
energies $\sim$$10^{60}$~erg. The original active galacic nucleus
(AGN) may be dead but the fossil will retain its stable
configuration resembling the reversed-field pinch (RFP)
encountered in laboratory MHD experiments.

In an RFP the longitudinal magnetic field changes direction at a
critical distance from the axis, leading to magnetic
re-connection there, and to slow decay of the large-scale RFP
field. We show that this field decay induces large-scale electric
fields which can accelerate cosmic rays with an $E^{-2}$
power-law up to ultra-high energies with a cut-off depending on
the fossil parameters.  The cut-off is expected to be rigidity
dependent, implying the observed composition would change from
light to heavy close to the cut-off if one or two nearby AGN
fossils dominate.  Given that several percent of the universe's
volume may house such slowly decaying structures, these fossils
may even re-energize ultra-high energy cosmic rays from
distant/old sources, offsetting the ``GZK-losses'' due to
interactions with photons of the cosmic microwave background
radiation and giving evidence of otherwise undetectable fossils.
In this case the composition would remain light to the highest
energies if distant sources or fossils dominated, but otherwise
would be mixed.  It is hoped the new generation of cosmic ray
experiments such as the Pierre Auger Observatory and ultra-high
energy neutrino telescopes such as ANITA and lunar Cherenkov
experiments will clarify this.
\end{abstract}

\begin{keywords}
acceleration of particles -- MHD -- magnetic fields -- galaxies: active -- galaxies: jets -- intergalactic medium 
\end{keywords}


\section{Introduction}

The energy spectrum of cosmic rays (CR) extends from below 1 GeV
up to at least $10^{20}$eV.  The bulk of the cosmic rays below
$\sim$$10^{18}$eV are probably Galactic.  It is currently unknown
whether the ultra-high energy (UHE) cosmic rays are Galactic or
extragalactic in origin, and whether they are accelerated or
result from emission by topological defects or decay of
super-massive particles.  If they are accelerated, they are
almost certainly extragalactic in origin.  This is because their
gyroradii are sufficiently large that an anisotropy would be
expected toward a galactic source population (not observed), and
because there are no obvious Galactic source populations having
magnetic fields such that gyroradii of UHE CR are compatible with
their size.  

Possible extragalactic acceleration sites include hotspots of
giant radio galaxies, the intergalactic medium, gamma ray bursts
and blazar jets.  Extragalactic CR at the highest energies are,
however, subject to interaction with the cosmic microwave
background radiation (CMBR).  In the rest frame of a UHE CR
proton, photons of the 2.73~K CMBR are strongly blue-shifted to
gamma-ray energies.  The thresholds for Bethe-Heitler pair
production and pion photo-production by UHE CR protons on the
CMBR are close to $2 \times 10^{17}$eV and $2 \times 10^{19}$eV,
respectively.  Protons at $3\times$$10^{19}$eV and
3$\times$$10^{20}$eV typically lose a large fraction of their
energy in a time of 1~Gpc/c ($3 \times 10^9$y) and 10~Mpc/c ($3
\times 10^7$y), respectively.  The importance of pion
photoproduction on the CMBR was first noted by \citet{Greisen66}
and \citet{Zatsepin66} and the cut-off they predicted is referred
to as the ``GZK cut-off''.  The pion photoproduction interactions
result in a flux of UHE neutrinos, often referred to as ``GZK
neutrinos'', which would point back to halos surrounding the UHE
cosmic ray sources.  Several experiments (\citet{Takeda03},
\citet{Bird95}, \citet{Connolly2006}, \citet{Abu-Zayyad05}) have
reported UHE CR events with energies above $10^{20}$~eV.  The
Auger array (\citet{Auger04}) will soon report its results with
excellent statistics.  Meanwhile, experiments such as ANITA
(\citet{ANITA-lite}) and lunar Cherenkov experiments are in
progress to detect the UHE neutrinos which should give clues to
the origin of UHECR.  See \citet{ProtheroeClay2004} and
\citet{SKAscienceCase} for recent reviews of UHECR and radio
techniques in UHE neutrino and particle astronomy.

Remnants of jets and their surrounding cocoons may persist long
after their parent AGN fade from view. These colossal MHD
structures decay slowly and yet may retain their relatively
stable self-organized configurations.  Decay depends on the
structure circuit resistance, and lifetimes could be quite long,
given the large inductance of the circuit, an initial outward
current along the jet and a return current back along an outer
sheath or cocoon around the jet.
The current closure condition requires that a large fraction of
magnetic fields resistively dissipate
\citep{Lesch1989A&A...225..341L}. Also
\citet{Lyutikov2007APh....27..473L} argue for direct inductive
acceleration in sheared relativistic jets while the jet are on,
though not afterwards. We concentrate here on the fossil era,
when a substantial fraction of jet energy has already been
invested in the magnetic energy of the fossil and induction
occurs through decay, not in the earlier building period --
though both eras can contribute to acceleration schemes.

Helical structures are common in AGN jets, arising during jet
formation from rotation of magnetized plasma accreting toward the
central black hole -- this could be enhanced in the case of
binary black hole systems.  Azimuthal electric currents are
therefore likely, yielding a magnetic field component along the
jet direction. 
In radio galaxy jets boundary layers are often clearly visible
in radio polarization (see,
e.g. \citet{LaingBridle2002MNRAS.336..328L} for observations and
interpretation of the jets of 3C~31).
\citet{Ostrowski2002MmSAI..73..387O} has argued that boundary
layers in jets are sites of particle acceleration.

Laboratory MHD experiments show that
reversed-field pinches are fairly stable. On the immense scale of
these ``fossil jets'', the decay time from instability can be
billions of years.  However, decay of these colossal MHD
structures on such time-scales result in electric fields capable
of accelerating existing populations of lower energy cosmic rays
up to ultra high energies with a flat spectrum extending from
some minimum rigidity (momentum/charge) determined by fossil
dimensions, magnetic field and decay time.  We first review the
properties expected of large magnetic fossils, based on extensive
work in the plasma physics community over the last four decades.

\section{Stability of MHD Structures}

We assume that the governing, evolved equilibrium of initially
jet-driven, then later, long-lived magnetic structures will be a
reversed field pinch (\citet{Taylor1986}, \citet{Bellan2000}). This axisymmetric configuration
has magnetic field components (in cylindrical coordinates -- $r,\phi,z$)
\begin{eqnarray}
B_z\sim B_{\phi}\sim \sqrt{\mu_0 P}
\end{eqnarray}
in an
MHD equilibrium made stable by optimally efficient radial
profiles, demanding a minimum of $B_z$ -- $B_{\phi}$ is built
by the jet current.

In a reversed-field pinch the crucial geometric feature is the
longitudinal magnetic field, which changes direction at some
critical distance from the jet axis. In experiments the RFP can be
curved into a torus, creating a spheromak (\citet{Bellan2000}). We
use a simple cylindrical RFP model below because, though a nearly
spherical geometry may be better for an evolved fossil
(especially in rich clusters with nearly isotropic pressures), the
spheromak geometry imposes many field curvatures effects in
particle orbits which are a complication beyond the essential
physical point of the reversed axial field.

Stable, low $\beta$ ($\beta=2\mu_0P/B^2$ where $P$ is the plasma pressure) reversed field pinch radial
profiles have several properties: 
\begin{enumerate} 
\item $B_z$
reverses near the outside of the confinement
region. This is the crucial shear that stabilizes modes that can
interchange flux tubes of plasma. It also prevents kink $(m=1)$,
current-driven instabilities.  
\item To suppress ``sausage" modes
driven by pressure, a value of $\beta < \frac{1}{2}$ is
essential. The smaller the $\beta$, the more stabilized a
confinement geometry becomes.  
\item A conducting ``wall" close
enough to the plasma core to suppress the kink, $m=1$,
current-driven internal kink modes. This also completely damps
all external kink modes.  
\item A pressure profile $P(r)$ that is
hollow or very flat in $r$, to suppress interchange of magnetic
flux tubes near the magnetic axis.  
\end{enumerate}

\citet{MelroseNichollsBroderick1994} showed that complicated
equilibrium models of an isolated, force-free magnetic flux tube
could be constructed with cylindrically symmetric, force-free
magnetic fields. These had a sequence of concentric layers with
piecewise-constant $a$, where $a$ is a running index of the
equilibria in concentric cylinders, each obeying the force-free
condition, $\vec{j}(r,\phi,z) = a(r)
\vec{B}(r,\phi,z)$. They used analytical methods to construct
models of the surface currents on these equilibria, finding some
conditions such that the net current flux in the tube was zero.

For example, a ``two-$a$ model'' permits current neutralization
with a non-zero magnetic flux. An infinite series of equilibria
are possible, yielding stable reversed-field geometries with
various boundary conditions. Often a nearby conducting boundary
simplifies the equilibrium as an ideal approximation.

	Are such equilibria at energy equipartition between
magnetic and relativistic plasma energies?  \citet{DunnFabian2004}
studied in cluster ``ghosts" the limits on $k/f$, where $k$ is the
ratio of the total relativistic particle energy to that in
electrons radiating between 10 MHz to 10 GHz, and $f$ is the
volume-filling factor of the relativistic plasma. Strikingly,
none of their ``bubbles'' had a simple equipartition between the
pressures from the relativistic particles and the magnetic
field. Furthermore, $k/f$ did not strongly depend on any physical
parameter of the host cluster. Though at first there seemed to be
two populations -- $k/f$ values around 2, another bunch around
300 -- this did not hold up (\citet{DunnFabianTaylor2005}). The
apparent bimodality of the $k/f$ distribution could be explained as
arising from two kinds of jets -- electron-positron, giving a low
value for $k$, and electron-proton. If protons are the extra
particles needed to maintain pressure equilibrium, but unseen in
the radio emission, $k$ is high. Also, bimodality could come from
either a non-uniform magnetic field, or a filamentary structure
in the lobes.

Both these possibilities are consistent with a reversed-field
equilibrium, since fields vary spatially. Variations in
re-acceleration, which may occur from reconnection events in the
field-reversal zone, could also affect the $k/f$ measure. At this
point we know too little to infer much. Clearly, though, constraints on the magnetic field of cluster bubbles found
by comparing the synchrotron cooling time to the bubble age show
that no bubbles in the sample are in equipartition.

This is plausible as laboratory experience with long-lived ``magnetic bottles"
such as tokomaks and reversed field pinches show that only those
far from equipartition survive long. Only when the plasma pressure
is far less by several orders of magnitude than magnetic field
energy density do structures long persist. This is an important
lesson of the applied fusion research program, now over half a
century old: only low $\beta$ is usefully stable.

In this spirit, we shall use the simplest possible
current-neutralized, reversed-field model, with a single alpha
and a nearby ideal conducting wall. It has zero total azimuthal
magnetic flux since some current is carried in our assumed
conducting wall. This simplification allows us to concentrate on
the particle acceleration mechanisms in a simple geometry.
Without this simplification, we would need a more complex
equilibrium.

\subsection{Evolution of Reversed-Field Magnetic Structures}

Recent detections of several ghost cavities in galaxy clusters
(\citet{Clarke06}) -- often, but not always, radio-emitting --
suggest that the cluster hot plasma stays well separated from the
bulk of the relativistic plasma on a timescale of $\sim$100
Myr. Some leakage is plausible, but robust survival does
occur. In cases where the radio-emitting cosmic-ray clouds are
confined within the observable hot gas reservoir of the cluster,
the dimensions of the X-ray hole combined with the calculated gas
pressure gives a ``laboratory-like'' measurement of the central
black hole's energy that is converted into $PdV$ work against the
intra-cluster medium (ICM). An example is the X-ray cluster MS
0735.6+7421 observed in X-rays (Chandra) by
\citet{McNamara2005Natur.433...45M} in which two large
$\sim$200~kpc cavities are seem in the X-ray emission; radio
emission comes only from the cavities.  The total energy from the
AGN outburst that created these giant magnetic bubbles, and
probably re-heated the cluster gas, is estimated to be $\sim 6
\times 10^{61}$ erg, of which a substantial fraction must be in
magnetic field based on pressure arguments.  Dynamical changes
will drive inductive electric fields (\citet{Ensslinetal1997},
\citet{Kronbergetal2004}).  Referring to galaxy cluster embedded
sources, \citet{Kronberg2005CosmicB} states that pressure balance
between the magnetic field and gas (thermal and CR gas) within
the holes, and the external thermal hot gas pressure implies
typical $B$ values of $\approx 20$--30~$\mu$G within the holes.

So, these ghosts may be RFPs. How have they survived so long?
\citet{Benford2006} proved a theorem: that for such jet-built RFP
structures, the same simple MHD stability conditions guarantee
stability, even after the jet turns off. This means that magnetic
structures made stable while a jet is on can evolve into fossils
that persist long after the building jet current has died
away. These may be the relic radio ``fossils'', ``ghost bubbles''
or ``magnetic balloons'' found in clusters and made visible by
contrast against the X-ray emission as seen in Hydra A by
\citet{Wiseetal06}. Giant radio galaxies such as Cyg A are
building such structures now. Such fossils have a massive
inventory of magnetic energy that can be $\sim$$10^{60}$ erg
(\citet{Kronbergetal2004}). Such enormous energy can only come
from the gravitational infall energy of a supermassive black
hole, when jets convey a few percent of the energy outward.

These equilibria will evolve under the magnetic tension that
sustains them. They are active, responsive structures, not a
disordered ball of magnetic fields. In rich clusters these
balloons may contract under external gas pressure. They can
later rise in the gravitational potential and then expand in the
lesser external pressure farther from the cluster core. Volume
change will alter the plasma pressure as $p\,dV$ work changes the
magnetic equilibrium. This in turn drives inductive electric
fields, which can accelerate ions. We envision such inductive
changes in response to external forces or internal evolution as
the deep cause of high energy particle acceleration.

The best examples of radio bubbles are 3C~84, M87 and Her~A.
The edges of these bubbles are amazingly sharp, even in projection,
as would be expected at the edge of a long-lived self ordered
magnetic structure.  Indeed, because of the helicity present in
the magnetic field of the original AGN jet, the magnetic bubble it
produces will retain this helicity and evolve to a self-organized
force-free structure such as an RFP or sheromak, which are hard
to pull apart -- the toroidal field lines opposing the the
lateral exansion of poloidal.  The magnetic fields of these
structures will therefore undergo minimal mixing with the overall
cluster magnetic field, giving rise to the sharp edges seen radio
images.  They should thus also have larger magnetic fields than the
surrounding intra-cluster medium.

Initially stable reversed field pinches become endangered by the
current-driven global kink instability. This seems to be the most
likely way for structures to fail when the jet current source is
on, since there is more free energy to drive instability. If a
structure survives the current-driven era, later stability seems
more probable, as loss of the jet removes free energy.  But
suppose the system fails to shed magnetic helicity $K$ as
dissipation of magnetic energy proceeds (by reconnection,
primarily). Such helicity loss is well known to lead to kink
instability. Here we can turn to abundant laboratory experience
as a guide.

\subsection{The Laboratory to Astrophysical Analogy}

Laboratory experiments provide much of our lore about reversed
field pinches. These are all low pressure devices (low $\beta$).
\citet{Taylor1963} argued that in all systems with $\beta <1$
magnetic reconnection conserves global magnetic helicity,
allowing stable geometries to evolve without losing stability
(see also \citet{Taylor1986}). This is so even without the
infinite conductivity assumption that is common in astrophysics.
``Unfreezing" the magnetic field lines from plasma demands
resistivity and thus some form of magnetic energy
dissipation. Even so, helicity can be preserved, as in the case
of dissipation by reconnection.  Thus $\beta$-values
approaching $1$ may occur in long-lived systems if helicity
(twist) may be shed through turbulent processes other than
reconnection.  Even
if the plasma $\beta$ in the jet is not small, after the jet
switches off and develops into as fossil jet with a
self-organized magnetic structure, plasma cooling and expansion
is expected to lead to lower $\beta$-values.  Recently, \citet{LapentaKronberg2005}
have compared detailed MHD simulations with radio observations of
the jet of 3C~303, a FR~II with a knotty structure, with the
knots appearing like self-contained magnetic structures launched
along the jet.  They find that in the knots $\beta \sim 10^{-5}$,
and that the jet is almost certainly elecromagnetically dominated
rather than plasma dominated.

A magnetically driven helical dynamo can also be described as
``dynamical magnetic relaxation". Magnetic relaxation describes
how structures in magnetic-pressure-dominated environments evolve
to their equilibrium states. The fully relaxed state is the
Taylor state, determined by minimizing the magnetic energy
subject to the constraint that magnetic helicity is
conserved. This leaves a force-free helical configuration with
the field relaxing to the largest scale available, subject to
boundary conditions. However, Taylor's theory by itself is not a
dynamical theory since it does not provide a time-dependent
description of how the large scale magnetic helicity evolves.

We can draw an analogy between spheromak or reversed-field pinch
formation and coronal field relaxation: the laboratory
configurations form because the injection of magnetic helicity
into a system takes the system away from the relaxed state.  But helicity injection
also drives tearing mode fluctuations that yield a finite,
turbulent electromotive force. This in turn drives the system to
a new relaxed state. In astrophysical coronae the situation is
similar. The helical field produced by a velocity-driven dynamo
inside a rotator gets buoyantly fed from below into the
corona. This takes the magnetically dominated corona away from
the relaxed state, but fluctuations arise, driving the system
back to a relaxed state. The corona thus acts like a ``laboratory"
plasma volume injected with helicity from below. Injection is not
at one site, but throughout the coronal-surface boundary. While
each of the individual injection sites (flux tubes) is like a
single laboratory configuration if the field twists get injected
along the tube, each tube can also open up to even larger scales
by interacting with other tubes. Magnetic energy then appears on
much larger scales -- a difference between the laboratory and
astrophysical cases.

The dynamo effect in laboratory plasmas is a self-organization
process. Resistive diffusion evolves the plasma away from a
preferred state. The dynamo forces the plasma back toward the
preferred, self-organized state.The resulting magnetic
configuration has excess free energy leading to instabilities (or
turbulence) that relax the plasma toward a state of lower
magnetic energy. The plasma relaxes to the lower energy state
through the ``dynamo" effect. The self-generation of plasma
current reconfigures the large-scale magnetic field to one with
lower energy.

The structure of the relaxed state is partly captured by
minimizing the magnetic energy in the plasma volume ($U_B =
\int(B^2/2\mu_0)dV$) subject to the constraint of constant
magnetic helicity ($K = \int \vec{A}\cdot\vec{B}\,dV )$, where
$\vec{A}$ is the magnetic vector potential). Magnetic helicity
topologically measures the knottedness of the magnetic field
lines. The minimization yields a magnetic field such that the
ratio of the current density to magnetic field, $j/B$, is a local
spatial constant (the Taylor state; \citet{Taylor1963}).
For a recent comprehensive discussion of these
effects, which subsumes much of the earlier discussion especially
as it unites theory and experiment, see \citet{Bellan2000}.

An experimental test of the Taylor conjecture inferred the change
in magnetic energy and helicity during a relaxation event
(\citet{Ji_etal1995}). An approximate measurement modeled the
instantaneous plasma state as a slowly varying MHD
equilibrium. Since the fields during relaxation change slower
than an Alfven time, the fields satisfy the MHD force balance
equation,
\begin{eqnarray}
\vec{\nabla}P= \vec{j}\times \vec{B}
\end{eqnarray}
through discrete field twisting and braiding
action. Solution of this equation with experimental constraints
found that during relaxation the magnetic energy reduces by about
8\%, while the magnetic helicity drops by about 3\%.

The Taylor conjecture gives a useful, experimentally verified framework to
depict approximately the final state of the relaxation, or
dynamo, process. But it provides no information on the physical
mechanism of the dynamo. Solution of the resistive nonlinear MHD
equations reveals the detailed dynamics. The spatial fluctuations
that underlie the dynamo are tearing instabilities driven by
spatial gradients in $j/B$, making the field lines tear and
reconnect. The MHD description of such spontaneous magnetic
reconnection has been investigated for several decades. Linear
theory shows that tearing instabilities grow on a timescale
between a short Alfven time ($L/v_A$), and the long resistive
diffusion time ($\mu_0L^2/\eta$, where $\eta$ is the resistivity). For
experimental plasma parameters, the theoretical growth time is
approximately 100~$\mu$s, comparable to the observed relaxation
time. In astrophysical jet-built structures, this can be billions
of years.

Tearing arises from a resonance between the fluctuating and mean
magnetic fields.  In laboratory plasma tearing occurs at many
radii within the plasma, leading to large-scale
reorganization. For stationary and homogeneous turbulence,
magnetic helicity $K$ is relatively conserved compared to magnetic
energy during relaxation. These dynamo effects play important
roles, conserving the total magnetic helicity, except for
resistive effects and a small battery effect.

In astrophysical magnetic structure relaxation, charge and current
imbalances adjust through the bulk, low energy plasma, which is
collisional. The higher energy population can take advantage of
the electric fields generated by decay, for they flow unimpeded
along field lines, driven steadily by the electric fields. This is the
essential picture in our model.

The turbulent dynamo effect converts magnetic helicity from the
turbulent magnetic field to the mean field, when the turbulence
is electromagnetic. When the turbulence is electrostatic or due
to the electron diamagnetic effect, the magnetic helicity of the
mean magnetic field gets transported across space. Two types of
electric fields get generated during reconnection in the RFP. One
is the helical electric field accompanying the helical tearing
mode that causes the reconnection. The second is a mean, or
axisymmetric, electric field induced by the flux change driven by
the tearing mode dynamo. Clearly this is a complicated
phenomenon.

We do not attempt to treat such details. Instead, we shall simple
take the inductive electric fields to be a fraction of the mean magnetic
field. In the absence of detailed knowledge of the structure, this seems a
good way to begin.

\subsection{Properties of a simple reversed field pinch}

\begin{figure}
\epsfig{file=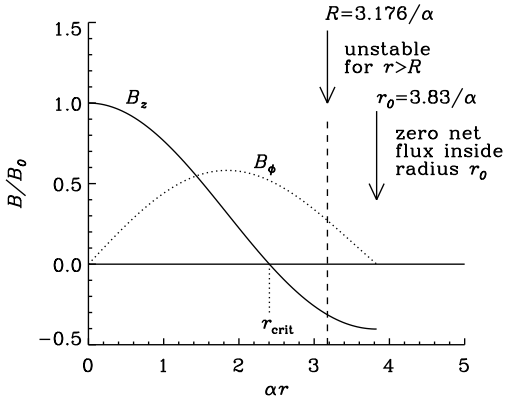, width=8cm}
\caption{Magnetic field components of an RFP.}
\label{Brfp}
\end{figure}

The simplest idealization would be for an
infinite cylindrical jet where the magnetic field in cylindrical
coordinates ($r,\phi,z$) is (\citet{Lundquist1951})
\begin{eqnarray}
B_r=0, \;\; B_\phi(r)=B_0 J_1(\alpha r), \;\; B_z(r)= B_0
J_0(\alpha r) ,
\label{lundquist}
\end{eqnarray}
and this has been shown (\citet{VoslamberCallebaut1962}) to be
stable for $\alpha r < 3.176$.  This radius for stability which
we take to be $R = 3.176/\alpha$ is where a conducting wall with
a large inertial mass would be present in an experimental
situation, and provide part of the circuit along which a return
current could flow, which we assume here to be the cocoon.  The
magnetic field is shown in Fig.~\ref{Brfp}.  Notice the
longitudinal field changes sign at $r_{\rm crit}=2.405/\alpha$,
the first zero of $J_0(\alpha r)$.

One can use Amp\`{e}re's law to find the current density
\begin{eqnarray}
\vec{j}(r) &=&  {1 \over \mu_0} (\vec{\nabla} \times \vec{B}) = {\vec{B}(r)\alpha \over \mu_0} \mbox{~~~(A m}^{-2}),
\end{eqnarray}
which is everywhere proportional to the magnetic field.
From this, the vector potential is 
\begin{eqnarray}
\vec{A}(r,\phi,z)={1\over \alpha }\vec{B}(r,\phi,z).
\label{vec_pot_equals_B}
\end{eqnarray}
The outward
electric current (positive $j_z$) is associated with radii $r < r_{\rm crit}$
and the return current with radii $r > r_{\rm crit}$,
\begin{eqnarray}
I_{\rm out} &\approx & (1.8\times 10^{18} \mbox{A})\times \left({B_0 \over {3 \;\mu {\rm G}}}\right) \left({R\over 100 \;{\rm kpc}}\right),\\ 
I_{\rm ret} &\approx & (5\times 10^{17} \mbox{A})\times \left({B_0 \over {3 \;\mu {\rm G}}}\right) \left({R\over 100 \;{\rm kpc}}\right)
\end{eqnarray}
and so some of the return current must be at radii $r>R$ --- in
the outer part of the cocoon or in the interstellar medium of the
host galaxy in order to exactly balance the outward
current.

The plasma pressure may be calculated from Eq.~1 of \citet{Robinson1971}, i.e.\
\begin{eqnarray}
{dP \over dr} + B_z{dB_z \over dr} + {B_\phi \over r}{d \over dr}(rB_\phi)=0.
\label{pgas}
\end{eqnarray}
Solving this for $\vec{B}$ given by Eq.~\ref{lundquist} results in
$P$=constant.  That is, for $\vec{B}$ given by
Eq.~\ref{lundquist} the tension of the azimuthal field lines
perfectly balances the radially outward pressure of the
longitudinal field.

\section{Charged particle motion and acceleration} 

The critical radius, $r_{\rm crit}$, where the
longitudinal field $B_z$ changes sign, is also the radius where
the longitudinal current density $j_z$ changes sign.  It is also
here that reconnection may occur and may be a site of particle acceleration.
\citet{LeschBirkSchopper2002} have considered acceleration in AGN jets due
to a reconnection electric field {\em along} the length of the
jet of an active quasar.  That paper shows acceleration of
hadrons to UHE should be possible in active AGN jets. The
{\em length} of the jet is critical in their model, determining
the maximum energy.  We differ in that the reconnection electric field
is azimuthal, i.e.\ {\em around} the fossil cylinder. Also, there should
be many nearby remnants, so they can accelerate UHECRs, whereas active quasars tend to be
at high redshift, and so their UHECRs will suffer the GZK cutoff if they are more than $\sim$10 Mpc away.

\subsection{Energy of Reconnection: Plasma Wave Model}

Rather than assign a fixed fraction of the ambient magnetic field
energy to the generated reconnection electric field, we can use
results from laboratory observations in reversed field pinches
(\citet{Ji_etal2004}; \citet{Ji_etal1995}). The zone where $B_z$ reverses will
probably be the reconnection site, since there only the azimuthal
field needs to reconnect.

In reconnection regions, micro-instabilities enhance resistivity,
driving the reconnection rate. These arise from both
electromagnetic and electrostatic modes. The electrostatic lower
hybrid drift instability has been invoked, but recent work shows
that it does not occur in the high-$\beta$ regions where
reconnection occurs. There, electromagnetic instabilities
prevail, as the $B_z$ field moves radially toward the
reconnection region, in our assumed cylindrical geometry,
preferentially annihilating in the zone where $B_z$ reverses
sign. This drives an azimuthal electric field. The plasma is
highly collisionless, so the induced resistivity greatly exceeds
the Spitzer level. The cause of this is ultimately the enhanced
modes, such as the modified two-stream instability, which in
experiment appear localized in the high-beta region where
reconnection occurs. Magnetic fluctuations $B_1$ appear also,
correlated strongly with resistivity enhancement (\citet{Ji_etal2004}).

We can estimate the equivalent electric field generated in
reconnection by noting that a momentum $p$ per unit volume
appears in the electromagnetic waves,
\begin{eqnarray}
p/V      =   {k\langle B_1^2\rangle \over \omega\mu_0}
\end{eqnarray}

Momentum moves from background plasma electrons and ions to the
electromagnetic fluctuations at a rate $2\gamma$, where $\gamma$ is the
nonlinear damping rate on the electron-ion background plasma. In
this highly nonlinear state, experiments show that $\gamma\sim\omega$, the
unstable mode frequency. Loss of momentum is a force equivalent
to an effective electric field in the reconnection region, $E_{\rm rec}$, given by
\begin{eqnarray}
enE_{\rm rec} = 2k\gamma {\langle B_1^2\rangle \over \omega\mu_0}
\end{eqnarray}

The magnetic wave fluctuation levels seen in experiment are of
magnitude $B_1 \sim 0.05$ the ambient averaged inflow field
levels, $B_z$. We thus set $B_1 = g B_z$ and the fraction $g=
0.05$.  Further, from experiment the observed wavelengths of this
turbulence are the size of the coherence lengths in the fields,
strongly suggesting a cooperative role (\citet{Ji_etal2004}). The
relevant wavenumber $k$ is the inverse Larmor radius of
background electrons in the plasma,
\begin{eqnarray}
r_L  = (10\mbox{ m})\times (T_{\rm eV})^{-1/2} \left({B_0\over \mu{\rm G}}\right)^{-1}
\end{eqnarray}
(\citet{Ji_etal2004}).

Scaling the results of laboratory experiments, we find 
\begin{eqnarray}
E_{\rm rec} &=& (1.5 \times 10^{-6} \mbox{  V m}^{-1})\times \left({g\over 0.05}\right)^2  \times\nonumber  \\
&&  \;\;\;\;  \times\left({B_z\over10 \mu{\rm G}}\right)^3(T_{\rm keV})^{1/2}\left({n\over 10^6\mbox{ m}^{-3}}\right)^{-1} 
\end{eqnarray}
with $n$ being the plasma density, and taking $B_z\sim 10 \,\mu$G
and assuming the heated region has keV energies.  To attain
fields of order $10^{-4}$ V/m then requires $g$ a bit larger, and
perhaps $B$ of 100 $\mu$G. Plasma temperature may also be higher
than keV, and density lower, also increasing $E_{\rm rec}$. 
These
parameter ranges would yield $10^{20}$ eV protons after several
encounters with such structures of scale $\sim$Mpc.  
Thus, we consider the possibility of UHE CR getting sequential
boosts in energy as a result of multiple encounters with
different fossils in a similar way that
\citet{IpAxford1992AIPC..264..400I} envisage Galactic cosmic rays
get re-energized by multiple interactions with supernova remnants
in the Galactic disk.

We expect it to be realistic to assume that electric
fields outside the reconnection zone can also accelerate
particles. The cold plasma (pressure is low) responsible for
currents which maintain the magnetic structure cannot short out
these electric fields, since they are inductively driven
everywhere in the structure, allowing acceleration
outside the reconnection zone.  Next we investigate whether or
not this alone will be viable for accelerating UHECR. In what
follows we assume that electric fields outside the reconnection
zone can accelerate particles.

\subsection{Global Electric Fields from RFP Decay}

Electric fields from
reconnection are {\em emf}s induced according to Faraday's law,
and so the electric field will be more extensive.  Assuming a
flow of flux lines toward the reconnection region, the
field will be changing everywhere and will
induce an electric field.  The simplest way of estimating
this global reconnection electric field is by assuming an
exponential decay of the magnetic field,
\begin{eqnarray}
\vec{B}(t) &=& B_0 e^{-t/t_{\rm dec}}[J_1(\alpha
r)\hat{\phi}+J_0(\alpha r)\hat{z}].
\end{eqnarray}
Then,
\begin{eqnarray}
\vec{E} &=& - {\partial \vec{A} \over \partial t}\;\; =\;\;
{\vec{B}\over t_{\rm dec}\alpha } \;,\\
\vec{E}  &\approx& 3.18\times 10^{-5} \left({B_0 \over
10\;\mu{\rm G}}\right)\left({R \over 100\;{\rm kpc}}\right) \left({t_{\rm
dec} \over {\rm Gyr}}\right)^{-1} \nonumber\\ && \times \,[J_1(\alpha
r)\hat{\phi}+J_0(\alpha r)\hat{z}]e^{-t/t_{\rm dec}}\mbox{~~~~(V/m)}
\label{ElecField}
\end{eqnarray}

\begin{figure}
\hspace*{-1cm}\epsfig{file=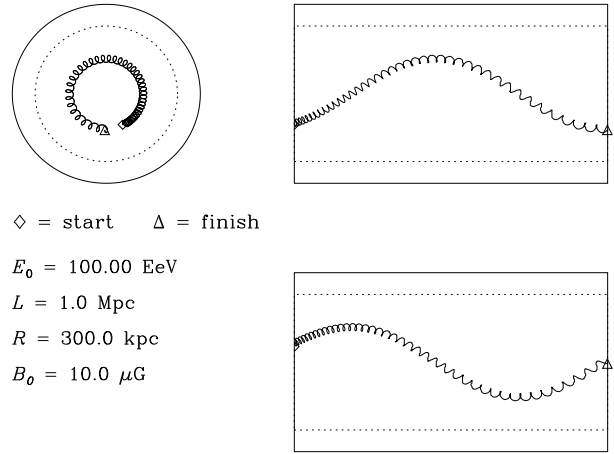, width=10cm}
\caption{Three orthogonal views showing a typical orbit in the RFP fields, and critical radius (dashed).}
\label{ExampleOrbits}
\end{figure}

The rate of change of the particle momentum $p=\beta\gamma mc$ (neglecting
losses) of a relativistic charge $Ze$ is then
\begin{eqnarray}
{d\vec{p} \over dt} &=& Ze c B (\vec{\beta}\times \hat{B} +\beta_{\rm dec} \hat{B}), \mbox{~~~where~~} \beta_{\rm dec} = {R
\over c t_{\rm dec}},
\label{dpdt}
\end{eqnarray}
such that a simple 1st order integration scheme might be
\begin{eqnarray}
{\vec{r}}(t+\Delta t) &=& {\vec{r}}(t)+c\Delta t \vec{\beta}(t),\\ 
\vec{p}(t+\Delta t) &=& \vec{p}(t)+{c\Delta t \over r_{\rm gyro}(\vec{r})}\, p \,\times \nonumber \\
&& \;\;\;\;\; \times
[\vec{\beta}(t)\times \hat{B}(\vec{r}) + \beta_{\rm dec} \hat{B}(\vec{r})],\\
\mbox{~~~where~~} r_{\rm gyro}(\vec{r}) &=& {\gamma \beta mc  \over Ze B(\vec{r})}.
\end{eqnarray}
Using this scheme we have conducted simulations of charged
particle motion in the RFP magnetic field including the effect of
energy change in the induced electric field.  We inject particles
uniformly and isotropically over the surfaces of disks of radius
$R$ at both ends of the jet, and follow their motion until they
escape.  A typical example is shown in Fig.~\ref{ExampleOrbits}.

\begin{figure}
\epsfig{file=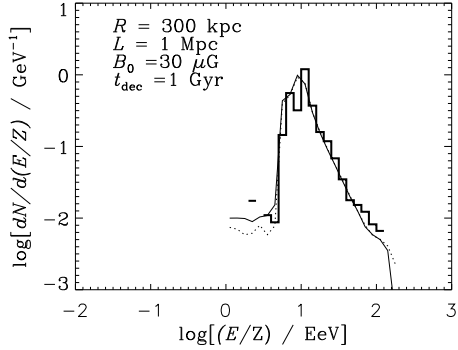, width=8cm}
\caption{Example output spectra (histograms) for monoenergetic
injection at $p_0c/Z=10^{18}$ eV, and fossil jet parameters as
specified and following particle trajectories as they undergo
helical motion along field lines.  Solid curve shows results for
case of neglecting gyromotion and simply following the gyrocentre
motion, and dashed curve shows effect of neglecting time
dependence of field (valid provided $t_{\rm max} \ll t_{\rm
dec}$).}
\label{ExampleSpectra}
\end{figure}

Ultra-relativistic particles of charge $Ze$ are injected with
energy $E_0$ (note for ultra-relativistic particles $E\approx pc$) and
their final energies are binned as shown in
Fig.~\ref{ExampleSpectra}.  Since the induced electric field is
in the same direction as the magnetic field, energy is gained as
particle move {\em along} field lines.  Thus we have also carried
out simulations in which the particle's position is simply
advanced along a field line, gaining momentum $ZeE\Delta t$ in
step $c\Delta t$.  Results for this case with high statistics
(solid curves) are in agreement with the results for full gyro
motion.  
We have shown that  for a decaying RFP, in which the
  electric field is everywhere parallel to the magnetic field,
  cosmic ray nuclei will gain energy unimpeded except by pion
  photoproduction on the CMBR -- as they undergo helical motion
  around magnetic field lines their momentum component parallel
  to $\vec{B}$ increases, while their momentum component
  perpendicular to $\vec{B}$ and the gyroradius remains constant,
  as is seen in the trajectory shown in Fig.~2.
Note that the ``double peaked'' structure (Fig.~3) is due to
separate contributions from injection at $r < r_{\rm crit}$ and
$r > r_{\rm crit}$, as we shall show in the next section.

\subsection{Analytic approach}

No approximations were made in obtaining the energy spectrum,
shown as the histogram in Fig.~3, by following particle
trajectories in the electromagnetic field of a decaying RFP.  To
obtain an approximate analytic result, we can assume that a
particle follows a helical path with its guiding centre on a
single field line from one end of the jet to the other.  Since the induced electric field is proportional to
the magnetic field according to Eq.~\ref{ElecField}, positively
charged particles will gain energy for pitch angles less than
$90^\circ$ and lose energy if their pitch angles are greater than
$90^\circ$.  

\begin{figure}
\epsfig{file=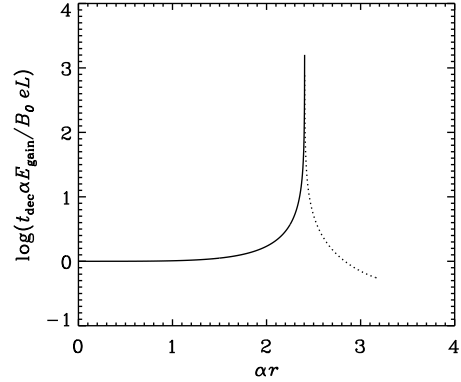, width=8cm}\\
\epsfig{file=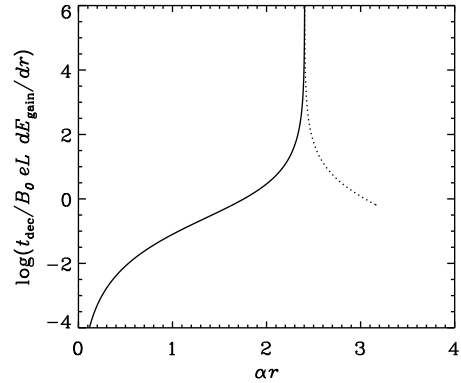, width=8cm}
\caption{(a) Momentum gain of particles injected at one end of the
RFP of length $L$ and exiting at the other.  Solid curve for positive particles traveling in positive $z$ direction, dotted for positive particles traveling in negative $z$ direction.  (b) The function $dp_{\rm gain}/ dr$}
\label{Brfp_accel}
\end{figure}

The energy gain on traversing the jet length $L$ will actually
depend on the pitch angle $\psi$ {\em of the helical magnetic
field line} acting as the guiding centre. So, positive particles
injected into the RFP with $r < r_{\rm crit}$ will gain energy
while moving in the positive $z$ direction, and those injected
with $ r_{\rm crit} < r < R$ will gain energy while moving in the
negative $z$ direction.  The increase in momentum of
ultra-relativistic particles of charge $Ze$ is
\begin{eqnarray}
p_{\rm gain} = {Ze E L\over c \,\cos \psi }
\end{eqnarray}
where 
\begin{eqnarray}
\vec{E} = {\vec{B}\over t_{\rm dec}\alpha } = B_0 {J_1(\alpha
r) \hat{\phi} + J_0(\alpha r) \hat{z}\over t_{\rm dec}\alpha },
\end{eqnarray}
\begin{eqnarray}
\cos\psi={B_z \over B}
\end{eqnarray}
Hence,
\begin{eqnarray}
p_{\rm gain}c &=& {Ze L\over t_{\rm dec}\alpha }{B^2\over B_z}=
{Ze B_0 L\over ct_{\rm dec}\alpha } \,{J_0(\alpha r)^2 +
J_1(\alpha r)^2 \over J_0(\alpha r)}, \label{eq:p_gain}\\
E_{\rm gain}  &=& E_{\rm gain}^0 \,{J_0(\alpha r)^2 +
J_1(\alpha r)^2 \over J_0(\alpha r)}
\label{p_gain}
\end{eqnarray}
where
\begin{eqnarray}
E_{\rm gain}^0 &\approx&   (10^{18}Z  \mbox{~eV~}) \left({B_0 \over 10\;\mu{\rm
G}}\right)\left({L \over {\rm Mpc}}\right) \times \nonumber \\ && \mbox{~~~~~~~} \times\left({R \over
100 \;{\rm kpc}}\right) \left({t_{\rm
dec} \over {\rm Gyr}}\right)^{-1}
\end{eqnarray}
and this is plotted in Fig.~\ref{Brfp_accel}(a).  Note that as $r \to r_{\rm crit}$, $p_{\rm gain} \to \infty$.

We can work out the energy spectrum as follows, 
\begin{eqnarray}
{dN \over dE_{\rm gain}} = {dN \over dp_{\rm gain}c} = {dN \over dr}\left[{dp_{\rm gain}c \over dr}\right]^{-1}
\label{dNdE}
\end{eqnarray}
where $dN/dr$ is the distribution in radius of the injection points.  For injection at one end of the RFP we would have uniform injection over the disk of radius $R$, giving
\begin{eqnarray}
{dN \over dr}={2r \over R^2} \mbox{  for  } 0 < r < R.
\end{eqnarray}
Differentiating Eq.~\ref{eq:p_gain} gives
\begin{eqnarray}
{dp_{\rm gain}c \over dr} = {Ze B_0 L\over t_{\rm dec} } \,\left[{J_1(\alpha r)J_0(\alpha r)^2- J_0(\alpha r)J_1(\alpha r)J_2(\alpha r) \over J_0(\alpha r)^2}\right]
\end{eqnarray}
which is plotted in Fig.~\ref{Brfp_accel}(b).

\begin{figure}
\epsfig{file=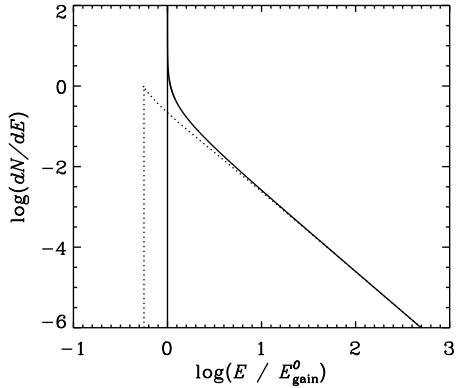, width=8cm}
\caption{Spectrum of accelerated particles.  (Curves have
same meaning as in Fig.~\protect\ref{Brfp_accel})}
\label{Brfp_spectra}
\end{figure}

\begin{figure}
\epsfig{file=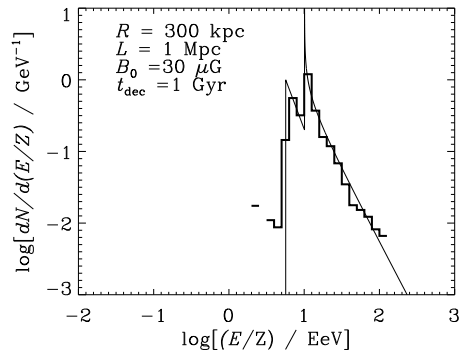, width=8cm}
\caption{Example output spectra (histograms) for monoenergetic
injection at $p_0c/Z=10^{18}$ eV, and fossil jet parameters as
specified and following particle trajectories as they undergo
helical motion along field lines.  Solid curve shows analytic
result.}
\label{ExampleSpectraComparison}
\end{figure}

Thus, from Eqn.~\ref{dNdE} we have ${dN / dp_{\rm gain}}$ as a
function of the parameter $r$, and from Eqn.~\ref{p_gain} we have
$p_{\rm gain}$ as a function of the parameter $r$, and so we can
plot ${dN / dp_{\rm gain}}$ vs.\ $p_{\rm gain}$, and this is
shown in in Fig.~\ref{Brfp_spectra}.  In
Fig.~\ref{ExampleSpectraComparison} we compare the analytical
spectrum with that obtained by following particle orbits.  This
analytic slope, which asymptotically is $E^{-2}$, is consistent
with that of Fig.~\ref{Brfp_spectra}.  The agreement reflects a
geometric property of the acceleration mechanism, as particles
near the reconnection region experience little axial magnetic
field. By our simplifying assumption that the inductive electric
field is proportional to the equilibrium magnetic field, we find
that such particles near the reconnection site then are
accelerated around the axis by the purely azimuthal electric
field. This generally keeps them in the structure longer, as they
do not move radially very much, so they leave only when they have
drifted down the axis to the end. Those which enter the fossil
near the reconnection zone will be preferentially accelerated,
becoming UHECRs. This gives a clean mapping of geometry into the
energy spectrum. Such spectra yield $10^{20}$ eV protons and iron
nuclei 26 times more energetic. Repeated accelerations in random
fossils along a particle route could maintain this quality if
fossils are common enough, and have sufficient decaying energy,
in our region of the universe. 

Finally, Heinz \& Sunyaev (2002) have
shown that particles should be accelerated at the reverse shock
of a micro-quasar jet colliding with the interstellar medium, and
that this may give a contribution to the galactic cosmic rays up
to about $\sim$10~GeV.  Here, we consider the possibility of
particle acceleration by electric fields in fossil micro-quasars
induced by the decay of a remnant magnetic bubble after the jets
have switched off, as in the case of fossil AGN.  This mechanism
might work for and pulsar wind nebulae as well as fossil
micro-quasars, and be responsible for emission in unidentified
TeV sources, as well as contributing to galactic CR.  In either
case the expected minimum energy would be
\begin{eqnarray}
E_{\rm gain}^0 &=&   (10^{12}Z \mbox{~eV~}) \left({B_0 \over 0.1\;{\rm
mG}}\right)\left({L \over 1 \,{\rm pc}}\right)  \times \nonumber \\
&& \times \left({R \over
1 \;{\rm pc}}\right) \left({t_{\rm
dec} \over {\rm Myr}}\right)^{-1} .
\end{eqnarray}

\section{Discussion}

Active galaxies have long been considered as a source
of the UHE CR, for example
\citet{Rachen1993A&A...272..161R} considered acceleration
at shocks associated with hot spots in lobes of Fanaroff-Riley II
(FR~II) radio galaxies.  There is no simple connection between
the maximum particle energy attainable in an AGN jet, in the hot
spot, and in the subsequent bubble.  In the case of a jet, either
the magnetic field through proton synchrotron radiation or pion
photoproduction interactions will stop the acceleration process
(\citet{MuckeProtheroe2001APh....15..121M}), whereas in the case
of hot spots in lobes of FR IIs the dimensions of the shock, its speed and
magnetic field configuration would determine the maximum energy.
Ultimately, pion photopruduction on the CMBR cannot be avoided,
and for shock acceleration with plausible parameters it is
difficult to exceed $\sim$$10^{21}$~eV
(\citet{Protheroe2004APh....21..415P}).  In the case FR~II jets
(or quasar jets) pion photoproduction on the photon field of a
luminous accretion disk will cut off the spectrum of accelerated
protons well below this.  However, in the case of FR I jets (or
the jets of BL Lac objects), having under-luminous disks or
advection dominated accretion flows, the cut-off will be due to
synchrotron losses or pion photoproduction on the jet synchrotron
photons.  The problem of escape of the highest energy protons
trapped by a jet's magnetic field can be solved by the same
photoproduction interactions, which convert protons to neutrons
which escape before decaying, and if travelling in the jet
direction the resulting cosmic ray protons have higher enrgies
from Doppler boosting (see \citet{Protheroe2003APh....19..559P}
for the case of M87).

Once accelerated to the ultra-high energy region, protons can
more readily escape their birth fossil. They will move through
the tens of Mpc where the GZK photo-pion scattering from the
background microwave radiation begins to take its toll. As well,
the UHECRs will scatter from other fossils. This can make them appear
isotropic if their directions are scattered enough, and they can
be further accelerated by these other fossils as well.

Our model is geometrically simple, to capture the essential
geometry of reverse field pinches, and the acceleration physics
that affords to particles, depending on where they are in the
structure. We have neglected important effects:
\begin{enumerate}
\item Scattering -- this in principle will rearrange particle
orbits, moving some out of the high-acceleration zone by
scattering their pitch angles, but also scattering others into
the preferential region.  In the highly ordered field of AGN
fossils, this may be less important ant for energization to UHE of
a pre-existing population of high energy cosmic rays accelerated,
e.g.\, in the host galaxy of the dead AGN.  Future work should
nevertheless include investigation of scattering of lower energy
particles by the magnetic turbulence expected in such sources,
generated by the inductive effects of their slow collapse and by
the ongoing reconnection events near the axial field reversal
sites.
\item Energy losses - here we have neglected energy losses by,
e.g.\ pion photoproduction and adiabatic expansion.  For the
range of model parameters used in the present work, these occur
more slowly than acceleration, but should be included in future
work.
\item Geometry. Assuming a cylindrical reversed field pinch
omits the many possible curvatures a more spherical RFP i.e., a
spheromak would entail. External pressure will not be isotropic,
generally. But if a fossil becomes detached from its parent
galaxy presumably by reconnection of fields at the feet of the
former jet structure it can rise, buoyant, in any cluster
gas. The shape will resemble a fat torus, with the currents
forming loops. We simplified this, feeling that torodial
curvature would complicate our calculations without adding new
physics. This can be relaxed in future, using more complicated
equilibria. 
\item Initial energy. To isolate the highest energy particle
issues, we began with a population of constant energy particles,
e.g.\ 1 EeV, but provided the injection energies are much less
than the minimum energy gained, by particles being accelerated
along the fossil jet, axis the final spectral shape is
insensitive to the initial energy.  Generalizing to an initial
energy spectrum will be done in future. This said, we have
produced spectra that yield very high energy particles. $10^{20}$
eV protons and iron nuclei 26 times more energetic still. If 1\%
of the universe is taken up with fossils (\citet{Kronberg1994}),
of size $<$ 1 Mpc, then there should be many fossils within 100
Mpc of us. A fresh structure is under construction at M87, 16 Mpc
away. This leads us to predict that while Auger should see a
smooth UHECR background above the GZK cutoff, $\sim$$6 \times
10^{19}$ eV. Auger may see clumps of UHECR in the sky, which
point back to nothing specially luminous. These will be fossils
we didn't know about--plus some we did. The UHECR will then
become a diagnostic of regions storing vast magnetic energies,
yet perceptible only through equally energetic particles,
bringing word of them across great distances.

\end{enumerate}

The total energy stored in the magnetic field of an RFP is
\begin{eqnarray}
U_B &=&1.3 \times 10^{59} \left({B_0 \over 10\;\mu{\rm
G}}\right)^2 \times  \nonumber \\
&& \;\;\;\; \;\;\;\; \;\;\;\;  \times \left({L \over {\rm Mpc}}\right)\left({R \over
100\;{\rm kpc}}\right)^2 \mbox{~~(erg)}.
\end{eqnarray}
These magnetic structures must have survived since the quasar
was last active, implying a decay time scale $t_{\rm dec}$ of the order
of Gyr, so
\begin{eqnarray}
\dot{U}_B &=& {U_B \over t_{\rm dec}} \; ,\\
\dot{U}_B  &=&  (4.0 \times 10^{42} \mbox{~~erg s$^{-1}$})\,\times \left({B_0 \over 10 \, \mu{\rm G}}\right)^2\,\times \nonumber\\
&& \;\;\;\; \times \,
\left({L \over 1 \, {\rm
Mpc}}\right)\left({R \over 100 \,{\rm kpc}}\right)^2 \left({t_{\rm dec}
\over {\rm Gyr}}\right)^{-1}.
\end{eqnarray}
assuming that the reconnection is continuous
and uniform along the cylindrical reconnection region, such that
the whole magnetic structure slowly decays exponentially on time scale
$t_{\rm dec}$ while maintaining the stable configuration given by
Eq.~\ref{lundquist}. 

\citet{ONeillJones_etal2006AN....327..535O} have examined the
interactions of AGN jets with magnetized cluster media using MHD
jet simulations in a cluster-like atmosphere extending over
distances $\sim$200 kpc.  They find that steady jets
asymptotically deposit $\sim$50\% of their power into ambient
thermal energy, $\sim$25\% into heating of the ICM, and the
remaining energy to be mostly stored in the jet backflow cocoon,
and that magnetic energy introduced by the jet is generally
amplified, especially by sheared flows; see also
\citet{Ryu1998A&A...335...19R}.

For an extragalactic source distribution locally producing an
$E^{-2}$ spectrum of protons, \citet{Lipari2005} estimates the local
power requirement to be $\sim 10^{50}$ erg Mpc$^{-3}$ y$^{-1}$.
Decaying magnetic fields with local filling
factor $\eta_{B}$ lose energy at a rate
\begin{eqnarray*}
\dot{u}_B & \sim &   10^{53} \eta_{B}\left({B_0 \over 10\;\mu{\rm
G}}\right)^2 \left({t_{\rm
dec} \over {\rm Gyr}}\right)^{-1} \mbox{~~~erg Mpc$^{-3}$ y~$^{-1}$}
\end{eqnarray*}
Magnetic fields from quasars can fill up to 5--20\% of the
intergalactic medium (\citet{FurlanettoLoeb2001}) -- probably higher
locally since our Galaxy is in a ``Wall''.  Indeed,
\citet{GopalKrishnaWiita2001ApJ...560L.115G} estimate the fractional relevant
volume that radio lobes born during the quasar era cumulatively
cover is $\sim$0.5.  Hence our crude
energetics arguments show fossil AGN structure decay could well be
responsible for the observed UHE CR.

The spectrum of accelerated particles will cut off at some
maximum momentum determined by either the finite thickness of the
reconnection zone (recall that in the analytic approximation as
$r \to r_{\rm crit}$, $p_{\rm gain} \to \infty$), or by the
gyroradius increasing so that it is no longer much less than the
radius of the fossil.  From Fig.~\ref{Brfp}, we see that for
$r<R$ the magnetic field is in the range $0.4B_0 <B < B_0$.
Hence, the condition $r_L \ll R$ implies
\begin{eqnarray}
E_{\rm gain}^{\rm max} &\ll& (10^{21}Z \mbox{~eV~}) \left({B_0 \over 10\;\mu{\rm
G}}\right)\left({R \over
100 \;{\rm kpc}}\right) .
\end{eqnarray}

The UHECR spectrum shows a steepening just below
$10^{18}$ eV (the "dip") followed by a flattening at around
$10^{19}$ eV (the "ankle") before the possible observed GZK
cut-off just above $10^{20}$ eV.
\citet{Berezinsky2006PhRvD..74d3005B} favour a scenario in which
extragalactic cosmic rays are injected with an $E^{-2.7}$
spectrum above above $\sim$$2\times10^{18}$ eV, and show that the
observed spectal feature referred to as the "dip" could then be
due to to Bethe-Heitler pair production on the CMBR.  While their
model fits are impressive, we believe that models in which the
galactic cosmic rays extend up to a few $10^{18}$ eV and have
extragalacic cosmic rays injected with an $E^{-2}$ spectrum, or
similar, e.g.\ following diffusive shock acceleration at
Fanaroff-Riley Class II radio galaxy hot-spots
\citep{Rachen1993A&A...272..161R}, are by no means ruled out.
Indeed, \citet{Hillas2006astro.ph..7109H} has shown that by
adding galactic and extragalactic spectra with various power-law
source spectra and compositions several possibilities, including
source spectra with an $E^{-2.2}$ spectrum, are consistent with
the data.  While a single ideal cylindrical RFP with particle
injection over its ends yields an $E^{-2}$ spectrum, the spectrum
due to several with different maximum energies would inevitably
be steeper.  Very recent results from the Pierre Auger
Observatory \citep{Yamamoto_arXiv:0707.2638} are not yet
conclusive, but appear to be consistent with source spectral
index in the range 2.2--2.5.

The spectrum of UHECR observed at Earth would have contributions
from nearby fossil jets at different distances, with different
powers and each having different dimensions and magnetic fields,
and hence a range of $E_{\rm gain}^0$ and $E_{\rm gain}^{\rm
max}$.  For an individual fossil, the cut-off is expected to be
rigidity dependent, implying the observed composition would
change from light to heavy close to the cut-off if one or two
nearby AGN fossils dominate.  However, if distant sources
dominate nuclei will be photo-disintegrated by interactions with
CMBR photons, and in this case the composition would remain light
to the highest energies if distant sources or fossils dominated.
Otherwise the composition could be mixed near the observed cut off.

Nearby potential sources of UHE CR would include the
bubbles around Cen~A, the nearest powerful radio galaxy, and around M87.
Radio observations at 20~cm of Cen~A by
\citet{Morganti1999MNRAS.307..750M} indicate the presence of
large features, one of which has been interpreted by
\citet{Saxton2001ApJ...563..103S} as a buoyant bubble of plasma
deposited by an intermittently active jet.  
Even closer, in M87 there are two bubbles of synchrotron emission
which appear to be inflated by outflows from the central engine
visible in the low frequency radio data of
\citet{Owen2000ApJ...543..611O}, with an amazingly sharp edge
indicating the possible presence of a shock. Given the very high
mass of the central black hole in M87, its jet, although fairly
powerful, is weak compared to its Eddington limit, making M87's
bubbles a prime cadidate as a UHE CR accelerator.

As we expect most of the fossil jets to be below the sensitivity
of current radio telescopes, it is impossible at the present time
to make firm predictions for the expected UHECR intensity at
Earth.  
A low synchrotron surface brightness, only observable if at all
at low frequency, does not necessarily mean that the magnetic
field is low.  Rather, we would argue that it is the cooling of
the relativistic electrons by synchrotron emission, since the
last AGN outburst which filled the bubble, that has reduced the
synchrotron emissivity.  However, such fossils may well observable when the SKA
(http://www.skatelescope.org/) is commissioned, hopefully within
$\sim$10 years.  
Nevertheless, we have demonstrated that it is
possible for this process to accelerate protons to UHE, and
nuclei to a $Z$ times higher energy, and shown that the power
requirements may reasonably be achieved given plausible volume
filling factors.

Very recent anisotropy data from the Pierre Auger Observatory
(\citet{Armengaud_arXiv:0706.2640},\citet{Harari_arXiv:0706.1715})
show no significant large-scale anisotropy or correlation of
cosmic ray arrival directions with BL~Lac objects.  Perhaps this
is not surprising because propagation of cosmic rays through
intergalactic magnetic fields will tend to isotropize their
arrival directions, except perhaps for UHE CR protons at the
highest energies from very nearby extragalactic sources.  With
longer exposure times Auger may see such anisotropies, in which
case the most probable source candidates would be bubbles around
Cen~A and M87.  Neurino observatories may, in principle, see UHE
neutrinos originating from directions close to UHE cosmic ray
acceleration sites, through pion photoproduction interactions of
the UHE CR on the CMBR during propagation from their sources.
These ``GZK neutrinos'' would appear as haloes around the true
source direction.  Of course, if such haloes appear where there
are no obvious source candidates, they could well be from fossil jets
with very low surface radio brightness.

Future work will include exploring observational consequences of
this model at radio, X-ray and gamma-ray energies.  Will electron
acceleration (in opposite direction to protons) be killed by
streaming instability, or by other processes?  If so, this will
heat the plasma.  Could this be seen in X-rays?  If not, radio
synchrotron and gamma-ray signals could be generated.  Could
these be detectable by existing radio telescopes or by the SKA in
coming decades, or by GLAST (http://glast.stanford.edu/)?  How
much of the energy of a fossil AGN goes into cosmic rays?
Finally, we mention the possibility that this mechanism operating
for microquasars might be responsible for acceleration of some of
the Galactic population of cosmic rays.

In conclusion, remnants of jets and their surrounding cocoons may
still be present around or close to galaxies which contain AGN
which are now no longer active.  These fossil jets are colossal
MHD structures and may have total energies $\sim 10^{60}$ erg.
After a jet switches off, the fossil jet will organize itself
into a stable configuration, such as the reverse-field pinch in
which the longitudinal magnetic field changes direction at a
critical distance from the axis, leading to re-connection there,
and to slow decay of the large-scale RFP field.  We have shown
that this field decay induces large-scale electric fields which
accelerate cosmic rays an $E^{-2}$ power-law up to ultra-high
energies.  Energetics arguments show that this provides a
plausible mechanism for the origin of the UHE CR.

\section*{Acknowledgents}

We thank Garang Yodh and Roger Blandford for stimulating
discussions, and KIPAC and SLAC for hospitality in May
2007.  We also thank the anonymous referee for helpful comments.  
RJP thanks the Australian Research Council for support through a
Discovery Project.

\end{document}